\documentclass[11pt]{article}

\usepackage[T1]{fontenc}
\usepackage[utf8]{inputenc}

\def\baselinestretch{1.28}
\global\arraycolsep=2pt

\usepackage[a4paper,top=30.6mm,bottom=38.6mm,left=34.6mm,right=34.6mm,footskip=1.3cm]{geometry}

\usepackage[nofancyfonts]{Packages}
\usepackage{slashed}
\allowdisplaybreaks

\acrodef{ybd}[\textsc{yb} deformation]{Yang--Baxter deformation}
\acrodef{nc}[\textsc{nc}]{non-commutativity}
\acrodef{sym}[\textsc{sym}]{super Yang--Mills}
\acrodef{himr}[\textsc{himr}  background]{Hashimoto--Itzhaki--Maldacena--Russo background~\cite{hep-th/9907166,hep-th/9908134}}
\acrodef{cybe}[\textsc{cybe}]{classical Yang--Baxter equation}
\acrodef{kl}[\textsc{kl}]{Kosmann--Lie}
\acrodef{sw}[\textsc{sw}]{Seiberg--Witten}
\acrodef{susy}[\textsc{susy}]{supersymmetry}

\DeclareMathOperator{\projop}{\Pi}
\newcommand*{\proj}[1]{\ensuremath{\projop^{\text{#1}}}}

\newcommand*{\setR}{\ensuremath{\mathbb{R}}}

\newcommand*{\del}{\mathop{\mathrm{{}\partial}}\mathopen{}}
\newcommand*{\tIIA}{type~\textsc{iia}\xspace}
\newcommand*{\tIIB}{type~\textsc{iib}\xspace}

\usepackage{hyperref}

\usepackage{cite}

\usepackage{fancyhdr}
\newcommand{\preprint}[1]{%
  \fancypagestyle{empty}{%
    \fancyhf{} %
    \fancyheadoffset[R]{\marginparsep+\marginparwidth}
    \fancyhead[R]{{#1}}
    \renewcommand{\headrulewidth}{0pt}
    \renewcommand{\footrulewidth}{0pt}
  }
}

\newcommand*{\AdS}[1]{\ensuremath{\mathrm{AdS}_{#1}}}

\newcounter{example}
\newenvironment{example}[2][]{\refstepcounter{example}\par\medskip
  \noindent \textbf{Example~\theexample. #1 #2}\\ \rmfamily}{\medskip\noindent}

\begin{document}

\numberwithin{equation}{section}

\preprint{KUNS-2742}

\begin{center}
{\Huge \textbf{SUSY and the bi-vector} }
\vspace*{1.5cm}\\
{\large 
Domenico~Orlando$^{\sharp,\flat}$, 
Susanne~Reffert$^{\flat}$, 
Yuta~Sekiguchi$^{\flat}$
and~Kentaroh~Yoshida$^{\ast}$
} 
\end{center}
\vspace*{0.2cm}
\begin{center}
  $^{\sharp}${\it INFN, sezione di Torino and Arnold--Regge Center\\
  via Pietro Giuria 1, 10125 Torino, Italy}
\vspace*{0.25cm}\\ 
$^{\flat}${\it Institute for Theoretical Physics, 
Albert Einstein Center for Fundamental Physics, 
University of Bern, Sidlerstrasse 5, CH-3012 Bern, 
Switzerland 
}
\vspace*{0.25cm}\\ 
$^{\ast}${\it Department of Physics, Kyoto University, \\ 
Kyoto 606-8502, Japan} 
\end{center}
\vspace{0.3cm}

\vfill

\begin{abstract}
  \noindent
  In this note we give an explicit formula for the preserved Killing spinors in deformed string theory backgrounds corresponding to integrable Yang--Baxter deformations realized via (sequences of) TsT transformations.
  The Killing spinors can be expressed only in terms of the bi-vector $\Theta$ which encodes the deformation.
  This formula is applicable to deformed backgrounds related to $r$-matrices of various ranks, including those that do not satisfy the unimodularity condition and give rise to backgrounds in generalized supergravity.
  We conjecture that our formula also remains valid for integrable deformations which are not realized via TsT transformations and motivate this conjecture by explicit examples.
\end{abstract}
\vfill

\setcounter{footnote}{0}
\setcounter{page}{0}
\thispagestyle{empty}

\newpage
\tableofcontents
\newpage
\section{Introduction} 
\label{sec:introduction}

The integrable structure behind the \AdS{}/CFT correspondence has lead to important progress as it allows the use of powerful integrability techniques (For a review, see \cite{arXiv:1012.3982}). Extending the reach of these techniques via integrable deformations of \AdS{}/CFT has been an important objective over the last years. A systematic way of generating integrable deformations are \acp{ybd}~\cite{hep-th/0210095,arXiv:0802.3518,arXiv:1308.3581,arXiv:1501.03665,arXiv:1309.5850,arXiv:1401.4855} which can be labeled by classical $r$-matrices satisfying 
the \ac{cybe}, or equivalently an anti-symmetric bi-vector $\Theta$ which corresponds to the non-commutativity parameter of the (generalized) Seiberg--Witten map~\cite{arXiv:1404.3657,arXiv:1504.05516,arXiv:1506.01023,arXiv:1610.05677,arXiv:1702.02861,arXiv:1705.02063,arXiv:1705.07116,arXiv:1708.03163,arXiv:1710.06849,arXiv:1803.05903,Hoare:2016wca,Hassler:2017yza,Lust:2018jsx,Bakhmatov:2018apn}.
\medskip

The issues of integrability and supersymmetry are quite disjoint. From the integrability point of view, whether or not any supersymmetries are preserved by an integrable deformation is often irrelevant. In fact, many string theory backgrounds stemming from integrable deformations do not preserve any supersymmetries at all.
From the point of view of deformed supersymmetric gauge theories realized via brane constructions in such deformed string theory backgrounds (the most prominent being the Omega deformation, see e.g.~\cite{arXiv:1106.0279,arXiv:1309.7350,Lambert:2018cht}), the amount of preserved supersymmetry and the explicit expressions for the preserved Killing spinors are on the other hand of crucial importance.

\medskip
In~\cite{Orlando:2018kms}, we gave a simple recipe for obtaining the explicit expression for the preserved Killing spinors in a class of deformed supergravity backgrounds. These backgrounds corresponded to integrable \acp{ybd}. We had focused in particular on \emph{unimodular} $r$-matrices leading to standard supergravity solutions and considered only $r$-matrices of rank 2 which could be realized by TsT transformations~\cite{arXiv:1404.1838,arXiv:1404.3657,arXiv:1412.3658, arXiv:1502.00740}.

\medskip
In this note we extend this result in a number of directions.
We give a \emph{frame-independent} formula for the preserved Killing spinors using the notion of the \ac{kl} derivative~\cite{Kosmann1971,Kelekci:2014ima}.
Moreover, we generalize our treatment to non-Abelian $r$-matrices. In~\cite{Borsato:2016ose}, it was shown by Borsato and Wulff that the only possible ranks for unimodular non-Abelian $r$-matrices satisfying the \ac{cybe} are rank 4 and rank 6.
While they give the full classification for rank 4, no systematic treatment of the rank-6 case exists. As however going to higher rank also translates into more broken supersymmetries, the rank-4 case is more interesting from a string/gauge theory point of view.

\medskip
We show here that our method applies equally to deformations corresponding to higher-rank $r$-matrices, which can be understood as resulting from sequences of TsT-transformations along non-commuting directions.
Our treatment moreover generalizes to \emph{non-unimodular} $r$-matrices giving rise to backgrounds obeying the \emph{generalized supergravity equations}~\cite{Arutyunov:2015mqj,Wulff:2016tju}. 

Even though our derivation relies on performing TsT transformations, the final result depends only one the antisymmetric bi-vector $\Theta$, which leads us to believe that it is valid beyond TsT-deformed backgrounds. We verify our formula for a number of cases which are \emph{not} realized via TsT transformations but are thought to result from non-Abelian T-dualities.
We therefore conjecture that our construction generalizes to any background deformation described by a generalized \ac{sw} map encoded by a bi-vector $\Theta$.

\medskip

Let us recall our previous results of~\cite{Orlando:2018kms} in order to introduce the notation.
A TsT-transformation consists of a combination of two T-dualities and a shift transformation. 
In a background with two isometry directions $\alpha_1$ and $\alpha_2$, a TsT transformation can be defined as
\begin{equation}
(\alpha_1, \alpha_2)_{\lambda} \equiv
    \begin{cases}
    \text{T-duality \(\alpha_1 \to \tilde \alpha_1\)} \\
    \text{shift \(\alpha_2 \to  \alpha_2 + \lambda \tilde \alpha_1 \)} \\
    \text{T-duality on \(\tilde \alpha_1\)} .
  \end{cases}
\end{equation}
The transformations $(\alpha_1, \alpha_2)_{\lambda}$ and $(\alpha_2, \alpha_1)_{\lambda}$ give the same result because there is another, geometric, commuting $SL(2,\setR)$ that permutes the two directions. 

An anti-symmetric bi-vector $\Theta$ can be introduced via the following field redefinitions 
(for the NS-NS sector)~\cite{Duff:1989tf,hep-th/9908142} that we read as a generalization of the \ac{sw} map:
\begin{equation}
  \label{eq:Theta}
  \begin{aligned}
    G_{mn} &= (g - B g^{-1}B)_{mn}, \\ 
    \Theta^{mn} &= -((g+B)^{-1}B(g-B)^{-1} )^{mn}, \\ 
    G_s &= g_s \left(\frac{\det(g+B)}{\det g}\right)^{1/2}. 
  \end{aligned}
\end{equation}
Here, $g_{mn}$, $B_{mn}$ and $g_s$ are the closed string metric, NS-NS two-form and closed string coupling, respectively. 
In analogy with the \ac{sw} map, we will refer to $G_{mn}$ and $G_s$ as the open string metric and coupling.

The bi-vector $\Theta$ is related to the isometry directions $\alpha_1$ and $\alpha_2$ associated to the transformation $(\alpha_1, \alpha_2)_{\lambda}$. It is given by (up to signs and numerical factors)~\cite{arXiv:1404.3657,arXiv:1504.05516,arXiv:1506.01023,arXiv:1610.05677,arXiv:1702.02861,arXiv:1705.02063,arXiv:1705.07116}
\begin{equation}
\begin{aligned}
\Theta = \frac{1}{2}\Theta^{\mu\nu}\partial_{\mu} \wedge \partial_{\nu} = \lambda \partial_{\alpha_1} \wedge \partial_{\alpha_2} .
\end{aligned}
\end{equation}
The last expression is nothing but a classical $r$-matrix expressed 
in terms of the Killing vector in the bulk.

The main result of~\cite{Orlando:2018kms} was a simple recipe for translating the {bi-vector $\Theta^{mn}$}
encoding the transformation data into an explicit formula for the preserved Killing spinors of the background resulting from the TsT transformation:\footnote{The exponential factor below is closely related to $\Omega$ in \cite{arXiv:1803.05903}. }
\begin{align}
  \label{eq:Killing-spinors-exp-proj}
  \epsilon_{-}^{\text{(fin)}} &= \proj{TsT} \epsilon_{-}^{\text{(in)}} , &
  \epsilon_{+}^{\text{(fin)}} &=  e^{\omega(\Theta) \, \frac{1}{2}\Theta^{\mu\nu}\hat{\Gamma}_{\mu\nu}}\proj{TsT} \epsilon^{\text{(in)}}_{+}\,,
\end{align}
where \(\epsilon^{\text{(in)}}\) are the Killing spinors in the initial background and \(\omega(\Theta)\) is a normalization factor satisfying
\begin{equation}
  \label{eq:omega_theta}
  \tan( \omega(\Theta) \sqrt{\frac{1}{2}\Theta^{mn}\Theta_{mn}}) = \sqrt{\frac{1}{2}\Theta^{mn}\Theta_{mn}}.
\end{equation} 
Here, we find that the formula for the Killing spinor is general and that the projector \(\proj{TsT}\) introduced in~\cite{Orlando:2018kms} is nothing but the solution to the algebraic equation
\begin{equation}\label{eq:projector}
  \bigg[\Theta^{\mu\nu}\Gamma_{\mu}\mathcal{S}\Gamma_{\nu} + \nabla_{\mu}\Theta^{\nu\rho}\Gamma^{\mu}{}_{\nu\rho} - 4\nabla_{\mu}\Theta^{\mu\nu}\Gamma_{\nu}\biggr] \proj{TsT} = 0\,.
\end{equation}
Note that it is completely determined by \(\Theta\). Note, that \(\nabla_{\mu}\Theta^{\nu\rho}\) is the non-geometric Q-flux \cite{arXiv:1710.06849}, and \(\nabla_{\mu}\Theta^{\mu\nu} = I^\nu\) is the extra Killing vector field appearing in generalized supergravity~\cite{arXiv:1702.02861,arXiv:1705.02063}.

We explicitly give the projectors and deformed Killing spinors of a number of deformed backgrounds which have already appeared in the literature in the context of integrable \acp{ybd}. We verify in particular the number of unbroken supersymmetries given in~\cite{Borsato:2016ose} for the examples taken from there.

\bigskip
The plan of this note is as follows. In Sec.~\ref{sec:Kosmann}, we introduce the notion of the \ac{kl}  derivative to discuss the supersymmetry variations in deformed backgrounds which also include the generalized supergravity case in a frame-independent way. In Sec.~\ref{sec:KosmannTheta}, we write the supersymmetry condition in a compact way depending only on the bi-vector $\Theta$. In Sec.~\ref{sec:rank-2}, we give the reformulation of some Abelian rank-2 examples which had already been treated in~\cite{Orlando:2018kms} as a warm-up. In Sec.~\ref{sec:higher-rank}, we give several explicit examples involving $r$-matrices of higher rank. In Sec.~\ref{sec:generalized-SUGRA}, give explicit examples corresponding to generalized supergravity backgrounds. In Sec.~\ref{sec:nonTsT}, we test our conjecture of the more general validity of our approach on three examples of rank-4 $r$-matrices which are not realized via TsT transformations. In Sec.~\ref{sec:conclusions}, we end with concluding remarks. In the Appendices, our conventions for the Gamma-matrices, the Killing spinors of the initial undeformed backgrounds and the expression for the complexified Killing spinor are collected.

\section{Killing spinors and the Kosmann-Lie derivative}
\label{sec:kosmann-lie}

\subsection{The Kosmann spinorial Lie derivative and Abelian T-duality}\label{sec:Kosmann}

Let us consider a ten-dimensional string background with a \(U(1)\) isometry.
In~\cite{Kelekci:2014ima} it was shown that the supersymmetries preserved by a T-duality in the direction of the isometry can be characterized in a frame-independent way in terms of the \ac{kl} derivative~\cite{Kosmann1971}.
Given a vector \(K = K^{m}\partial_{m}\) and a spinor \(\epsilon\), we define the \ac{kl} derivative as~
\begin{equation}
  \label{eq:zeroKosmann}
  \mathcal{L}_{K}\epsilon = K^{m}\nabla_{m}\epsilon + \frac{1}{4}(\nabla K)_{mn}\Gamma^{mn}\epsilon\,,
\end{equation}
where the $\nabla_{m} = \partial_{m} + \frac{1}{4}\omega_{m}^{ab}\Gamma_{ab}$.
It was shown in~\cite{Kelekci:2014ima} that after a T-duality in the direction of \(K\), only the supersymmetries satisfying \(\mathcal{L}_{K} \epsilon = 0\) are preserved.

Here we would like to extend this result to include the case of generalized supergravity.
We can always choose a frame in which \(K = \del_z\) and the metric takes the form
\begin{equation}
  \dd{s}^{2} = g_{\mu\nu}(x) \dd{x^{\mu}} \dd{x^{\nu}} + e^{2C(x)} \left(\dd{z} + A_{1}(x)\right)^{2}\, .
\end{equation}
In this frame, the Lie derivative on the spinor is simply
\begin{equation}
  \mathcal{L}_{\partial_{z}} \epsilon = \partial_{z}\epsilon = 0\,.
\end{equation}
Let us start from a generalized type IIA background of the form
\begin{equation}
  \begin{aligned}
    ds^{2} &= g_{\mu\nu} dx^{\mu}dx^{\nu} + e^{2C}(dz+A_{1})^{2}\,,\\
    B_{2} &=B + B_{1} \wedge dz\,,\\
    \mathfrak{f}_{0} &= m\,,\\
    \mathfrak{f}_{2} &= \mathfrak{g}_{2} + \mathfrak{g}_{1} \wedge (dz+A_{1})\,,\\
    \mathfrak{f}_{4} &= \mathfrak{g}_{4} + \mathfrak{g}_{3} \wedge (dz+A_{1})\,,\\
    \Phi &= -a z + \varphi + \frac{1}{2}C\,,\\
    I &= \hat{a}\,\partial_{z}\,,
  \end{aligned}
\end{equation}
where $A_{1}\,, B\,, B_{1}\,,\mathfrak{g}_{k}, \varphi, C$ depend only on $x^{\mu}$\,.

Even though the dependence of the dilaton on $az+\varphi$ in principle precludes us from T-dualizing in $z$, we can still formally perform a generalized T-duality along $z$, yielding a generalized type IIB background (note in particular the non-standard transformation of the dilaton):
\begin{equation}
  \begin{aligned}
    d\tilde{s}^{2} &= g_{\mu\nu} dx^{\mu}dx^{\nu} + e^{-2C}(dz+B_{1})^{2}\,,\\
    \tilde{B}_{2} &=B + B_{1} \wedge dz\,,\\
    \mathfrak{f}_{1} &= m (dz+B_{1}) - \mathfrak{g}_{1}\,,\\
    \mathfrak{f}_{3} &= \mathfrak{g}_{2} \wedge (dz+B_{1}) - \mathfrak{g}_{3} \,,\\
    \mathfrak{f}_{5} &= (1+ \star_{10})\bigl[\mathfrak{g}_{4} \wedge (dz+B_{1})\bigr]\,,\\
    \tilde{\Phi} &= - \hat{a} z + \varphi - \frac{1}{2}C\,,\\
    I &= a\,\partial_{z}\,.
\end{aligned}
\end{equation}
The background fluxes $\mathfrak{f}_{k}$ are related to the generalized fluxes via $\mathfrak{f}_{k} = e^{-C/2}\mathcal{F}_{k}$ in type IIA and $\mathfrak{f}_{k} = e^{C/2}\mathcal{F}_{k}$ in \tIIB.
We use \(z\) for both the coordinate in \tIIA and its dual in \tIIB.

Consider now the supersymmetry variations in the two frames above.
In generalized type IIA supergravity, the variation of the dilatino and gravitino read respectively,
\begin{equation}
\begin{aligned}
\delta \lambda &= \frac{1}{2}\slashed{\partial}\Phi \epsilon + \frac{1}{2}(I^{m}B_{mn} +I_{n}\sigma^{3})\Gamma^{n}\epsilon - \frac{1}{24}\slashed{H}_{3}\sigma^{3}\epsilon + \frac{1}{8}\biggl[5\mathcal{F}_{0} \sigma^{1} + \frac{3}{2}\slashed{\mathcal{F}}_{2}(i\sigma_{2})+ \frac{1}{24}\slashed{\mathcal{F}}_{4}\sigma^{1}\biggr]\epsilon\,,\\
\delta \Psi_{m} &= \nabla_{m} \epsilon - \frac{1}{8}H_{3mnp}\Gamma^{np}\sigma^{3}\epsilon + \frac{1}{8}\biggl[\mathcal{F}_{0} \sigma^{1} + \frac{1}{2}\slashed{\mathcal{F}}_{2}(i\sigma^{2})+ \frac{1}{24}\slashed{\mathcal{F}}_{4}\sigma^{1}\biggr]\Gamma_{m}\epsilon,
\end{aligned}
\end{equation} 
whereas in generalized type IIB,
\begin{equation}
\begin{aligned}
\delta \tilde{\lambda} &= \frac{1}{2}\slashed{\partial}\tilde{\Phi} \tilde{\epsilon} + \frac{1}{2}(\tilde{I}^{m}\tilde{B}_{mn} +\tilde{I}_{n}\sigma^{3})\Gamma^{n}\tilde{\epsilon} - \frac{1}{24}\slashed{\tilde{H}}_{3}\sigma^{3}\tilde{\epsilon} + \frac{1}{8}\biggl[\mathcal{F}_{1} (i\sigma^{2}) + \frac{1}{12}\slashed{\mathcal{F}}_{3}\sigma_{1}\biggr]\tilde{\epsilon}\,,\\
\delta \tilde{\Psi}_{m} &= \nabla_{m} \tilde{\epsilon} - \frac{1}{8}\tilde{H}_{3mnp}\Gamma^{np}\sigma^{3}\tilde{\epsilon} - \frac{1}{8}\biggl[\mathcal{F}_{1} (i\sigma^{2}) + \frac{1}{6}\slashed{\mathcal{F}}_{3}\sigma^{1}+ \frac{1}{240}\slashed{\mathcal{F}}_{5}(i\sigma^{2})\biggr]\Gamma_{m}\tilde{\epsilon},
\end{aligned}
\end{equation} 
where the Killing spinor is expressed as the doublet
\begin{equation}
\epsilon = \begin{pmatrix}\epsilon_{+}\\\epsilon_{-}\end{pmatrix}.
\end{equation}
Substituting the above background data into the variations, we see that the two sets of variations are related by
\begin{align}
  \delta\tilde{\Psi}_{\mu +} &= -\Gamma_{{\underline z}}\delta \Psi_{\mu + }\,, & \delta \tilde{\Psi}_{\mu -} &= \delta \Psi_{\mu - } \,,
\end{align}
if the \ac{kl} derivative along \(z\) vanishes,
\begin{equation}
  \del_z \epsilon_{\pm} = 0,
\end{equation}
and
\begin{align}
  \tilde \epsilon_{+} &= -\Gamma^{\underline{z}}\epsilon_{+} \,,&
                                                                  \tilde \epsilon_{-} &= \epsilon_{-}\,,
\end{align}
where $\Gamma_{{\underline z}}$ and $\Gamma^{{\underline z}}$  are the flat Gamma matrices.  \\
The final result is that a Killing spinor \(\epsilon\) in \tIIA is mapped to a Killing spinor \(\tilde \epsilon\) in \tIIB if and only if its \ac{kl} derivative along the isometry vanishes.
In other words, a supersymmetry is preserved if and only if it is \emph{independent} of the isometry direction (in the sense of the \ac{kl} derivative).\footnote{This conclusion is the same as in standard supergravity because the extra term proportional to \(I_n\) in the dilatino equation compensates the extra term in the transformation of the dilaton under generalized T-duality.}

\subsection{The Kosmann Lie derivative and the bi-vector $\Theta$}\label{sec:KosmannTheta}

It has been shown recently that if we perform the field redefinitions given in Eq.~(\ref{eq:Theta}) for 
\textsc{yb}-deformed backgrounds fulfilling the homogeneous \ac{cybe}, the open string metric $G_{mn}$ becomes the original undeformed metric, the open string coupling becomes constant and all of the information of the deformation is encoded in the bi-vector $\Theta^{mn}$~\cite{arXiv:1506.01023, arXiv:1610.05677,  arXiv:1702.02861,  arXiv:1705.02063, arXiv:1705.07116}. 
This means that the bi-vector $\Theta^{\mu\nu}$ introduced in Eq.~\eqref{eq:Theta} is equal to the classical $r$-matrix. 

In the case of a \textsc{yb}-deformation, 
we have
\begin{equation}
  \Theta^{\mu\nu} = r^{\mu\nu}.
\end{equation}
More generally, we can incorporate the classical $r$-matrix into the non-commutative bi-vector using the bi-Killing structure:
\begin{equation}
  \begin{aligned}
    \Theta &= \frac{1}{2}\Theta^{\mu\nu}\partial_{\mu}\wedge \partial_{\nu}\\
    &= \frac{1}{2}r^{ij}K^{\mu}_{i}K^{\nu}_{j}\partial_{\mu}\wedge  \partial_{\nu}\,,
  \end{aligned}
\end{equation}
where the Killing vectors $K_{i}$  form an algebra with structure constants $c_{ij}^{k}$:
\begin{equation}
[K_{i}, K_{j}] = c_{ij}^{k}K_{k}.
\end{equation}

In the commutative case, the deformation of the bulk geometry is obtained as a TsT transformation in the directions \(\del_\mu\) and \(\del_\nu\).
Here we will show that in this case, the condition for the preservation of supersymmetry can be completely expressed in terms of \(\Theta\) in a frame-independent fashion.

Let \(\epsilon \) be a Killing spinor for the (initial) undeformed background.
Then the gravitino variation vanishes:
\begin{equation}
  \label{eq:undeformed-gravitino}
  \delta \Psi_{\mu} = \nabla_{\mu}\epsilon + \frac{1}{8}\mathcal{S}\hat{\Gamma}_{\mu}\epsilon=0\,,
\end{equation}
where $\mathcal{S}$ is the analog of the Ramond--Ramond flux bispinor in the undeformed background:
\begin{equation}
  \mathcal{S} = -\slashed{\mathcal{F}}_{1}\otimes(i\sigma_{2}) - \frac{1}{3!}\slashed{\mathcal{F}}_{3}\otimes \sigma_{1}-\frac{1}{2\cdot5!}\slashed{\mathcal{F}}_{5}\otimes (i\sigma_{2}).
\end{equation}
In the cases of flat space or AdS${}_5 \times S^5$, it takes a particularly simple form:
\begin{align}
  \mathcal{S}_{\mathrm{o}} &= 0 & \text{flat space} ,\\
  \mathcal{S}_{\mathrm{o}} & = -\frac{1}{2\cdot 5!} \slashed{\mathcal{F}}_{5}\otimes (i\sigma_{2}) =  -2 \pqty{ \slashed{\omega}_{\AdS{5}} + \slashed{\omega}_{S^5}} \otimes (i\sigma_{2})& \text{\(\AdS{5} \times S^5\),} 
\end{align}
where \(\omega\) is the volume form.
As discussed above, the supersymmetries that survive a TsT transformation are those for which the \ac{kl} derivative along the two Killing spinors vanishes:
\begin{equation}
  \label{eq:vanishing-Kosmann}
  \mathcal{L}_{K_{i}}\epsilon = K^{\mu}_{i}\nabla_{\mu}\epsilon + \frac{1}{4}(\nabla K_{i})_{\mu\nu}\Gamma^{\mu\nu}\epsilon =0\,.
\end{equation}
Now, projecting the gravitino equation on \(K\) we find
\begin{equation}
  K^{\mu}_{i}\delta\Psi_{\mu} =K^{\mu}_i\nabla_{\mu}\epsilon + \frac{1}{8}K^{\mu}_{i}\mathcal{S}\Gamma_{\mu}\epsilon=0\,,
\end{equation}
which we can use to remove the covariant derivative on the spinor $\nabla_{\mu}\epsilon$ in the \ac{kl} equation and obtain the condition
\begin{equation}
  \biggl[\frac{1}{8}K^{\mu}_{i}\mathcal{S}\Gamma_{\mu} - \frac{1}{4}(\nabla K_{i})_{\mu\nu}\Gamma^{\mu\nu}\biggr]\epsilon = 0\,.
\end{equation}
Multiplying by $r^{ij}K^{\rho}_{j} \Gamma_{\rho}$ we get
\begin{equation}
  \biggl[\frac{1}{8}r^{ij}K_{j}^{\rho}\Gamma_{\rho}K^{\mu}_{i}\mathcal{S}\Gamma_{\mu} - \frac{1}{4}r^{ij}K_{j}^{\rho}(\nabla K_{i})_{\mu\nu}\Gamma_{\rho}\Gamma^{\mu\nu}\biggr]\epsilon =0\,,
\end{equation}
where we recognize the covariant derivative of \(\Theta\):
\begin{equation}
  \nabla_{\mu}\Theta^{\nu\rho} = 2 r^{ij}(\nabla_{\mu}K^{[\nu}_{i})K^{\rho]}_{j} \, ,
\end{equation}
so that the condition becomes
\begin{equation}
  \quad  \biggl[-\frac{1}{8}\Theta^{\rho\mu}\Gamma_{\rho}\mathcal{S}\Gamma_{\mu} - \frac{1}{4}r^{ij}K_{j}^{\rho}(\nabla K_{i})_{\mu\nu}\Gamma_{\rho}{}^{\mu\nu} + \frac{1}{2}r^{ij}K_{i}^{\rho}\nabla_{\rho} K_{j\nu}\Gamma^{\nu}\biggr]\epsilon =0 .
\end{equation}
Using the Killing vector equation  $\nabla_{\{\mu}K_{\nu\}} = 0$, we obtain the final form of the supersymmetry conditions:
\begin{equation}
  \label{eq:NCeq}
  \begin{cases}
    \nabla_\mu \epsilon + \frac{1}{8} \mathcal{S} \hat \Gamma_\mu \epsilon = 0 ,\\
    \bigg[\Theta^{\mu\nu}\Gamma_{\mu}\mathcal{S}\Gamma_{\nu}+\nabla_{\mu}\Theta^{\nu\rho}\Gamma\indices{^{\mu}_{\nu\rho}} - 4\nabla_{\mu}\Theta^{\mu\nu}\Gamma_{\nu}\biggr]\epsilon = 0\,,
  \end{cases}
\end{equation}
where we stress once more that the gravitino variation must vanish in the undeformed (open string) background. This formula only depends on $\Theta^{\mu\nu}$, so it is natural to conjecture that it can be applied to any background deformation which is labeled by the bi-vector $\Theta$, even though the proof given above is true only for deformations which are (sequences of) TsT transformations.

An important caveat is that in passing from the equation for a single Killing vector \(K\) to the one for the bi-vector \(\Theta\) we have implicitly assumed that \(K\) is not light-like.
If the bi-vector can be decomposed as \(\Theta = K_1 \wedge K_2 \) where \(K_1\) is light-like, our formula only gives a necessary condition for the preservation of supersymmetry.

\bigskip
The projector \(\proj{TsT}\) introduced in~\cite{Orlando:2018kms} is nothing else than the solution to the algebraic equation
\begin{equation}\label{eq:projector}
  \bigg[\Theta^{\mu\nu}\Gamma_{\mu}\mathcal{S}\Gamma_{\nu} + \nabla_{\mu}\Theta^{\nu\rho}\Gamma^{\mu}{}_{\nu\rho} - 4\nabla_{\mu}\Theta^{\mu\nu}\Gamma_{\nu}\biggr] \proj{TsT} = 0\,.
\end{equation}
The explicit form of the conserved Killing spinors in the deformed background is
\begin{align}
  \label{eq:Killing-spinors-exp-proj}
  \epsilon_{-}^{\text{(fin)}} &= \proj{TsT} \epsilon_{-}^{\text{(in)}} , &
  \epsilon_{+}^{\text{(fin)}} &=  e^{\omega(\Theta) \, \frac{1}{2}\Theta^{\mu\nu}\hat{\Gamma}_{\mu\nu}}\proj{TsT} \epsilon^{\text{(in)}}_{+}\,,
\end{align}
where \(\epsilon^{\text{(in)}}\) are the Killing spinors in the undeformed background.
This expression for the Killing spinor was derived for $r$-matrices of rank 2, corresponding to a single TsT transformation. It is straightforward to see that it generalizes to the case of higher rank. We conjecture that it remains valid also in general.

\bigskip
Note that the second and the third term in the projector equation~\eqref{eq:projector} have a natural interpretation in terms of generalized supergravity: \(\nabla_{\mu}\Theta^{\nu\rho}\) is the non-geometric Q-flux \cite{arXiv:1710.06849}, and \(\nabla_{\mu}\Theta^{\mu\nu} = I^\nu\) is the extra Killing vector field~\cite{arXiv:1702.02861,arXiv:1705.02063}.
We can rewrite the supersymmetry equation in the compact form
\begin{equation}
    \bigg[\Theta^{\mu\nu}\Gamma_{\mu}\mathcal{S}\Gamma_{\nu} + Q\indices{_\mu^{\nu \rho}} \Gamma\indices{^{\mu}_{\nu\rho}} - 4 I^\mu \Gamma_\mu \biggr]\epsilon = 0.
\end{equation}

In the following we will consider some explicit examples of applications of the projector equation \eqref{eq:NCeq}.
First we reproduce some results from~\cite{Orlando:2018kms} and then we move on to more general deformed backgrounds.

\section{Killing spinors for unimodular $r$-matrices of rank two}
\label{sec:rank-2}

As a warm-up, we revisit two examples we discussed in~\cite{Orlando:2018kms} which we now re-cast in a frame-independent way.
\begin{example}{$r=P_{1}\wedge P_{2}$}
First, we consider the classical $r$-matrix associated with the TsT-transformation $(x^{1},x^{2})_{\lambda}$ on $\AdS{5}\times S_{5}$. In this case, the bi-vector \cite{arXiv:1404.3657}
\begin{equation}
\Theta = \lambda \partial_{1} \wedge \partial_{2}
\end{equation}
is obviously divergence-free: $\nabla_{\mu}\Theta^{\mu\nu} = 0\,.$ Therefore, the left-hand side of the projector equation consists of two relevant contributions:
\begin{equation}
\begin{aligned}
\left[\Theta^{\mu\nu}\Gamma_{\mu}\mathcal{S}\Gamma_{\nu}  + \nabla_{\mu}\Theta_{\nu\rho}\Gamma^{\mu\nu\rho}\right]\epsilon %
&\sim \left[1+ \gamma \Gamma_{z} \otimes (i\sigma_{2})\right]\epsilon =0\,,
\end{aligned}
\end{equation}
where $\gamma = \Gamma_{\theta_{1}\theta_{2}\phi_{1}\phi_{2}\phi_{3}} = \Gamma_{56789}$ and $\Gamma_{012\cdot\cdot\cdot789}\epsilon =\epsilon$.
This implies that the non-zero solution for the Killing spinor equations is subject to the projection condition given by
\begin{equation}
(1 + \gamma \Gamma_{z} \otimes (i\sigma_{2}))\epsilon=0\,.
\end{equation}
Therefore, we conclude that the background deformed by the TsT transformation $(x^{1},x^{2})_{\lambda}$ preserves 16 supercharges due to the insertion of the projector
\begin{equation}
\Pi^{\mathrm{TsT}} = \frac{1}{2}(1+ i\gamma \Gamma_{z})
\end{equation}
in front of the constant complex spinor.
\end{example}

\begin{example}{$r=(M_{23}+M_{45})\wedge (M_{45}+M_{67})$}
The special virtue of equation \eqref{eq:NCeq} is that it is possible to extract all the projectors from this single equation. We first study a TsT transformation on flat space which mimics the Lunin--Maldacena deformation in~\cite{hep-th/0502086} (see also~\cite{Matsumoto:2015ypa} for integrable deformations of flat spacetime).
  The corresponding bi-vector is given by \cite{arXiv:1404.1838}
\begin{equation}
\Theta = \lambda (\partial_{\phi_{1}} \wedge \partial_{\phi_{2}} + \partial_{\phi_{2}} \wedge \partial_{\phi_{3}} + \partial_{\phi_{3}} \wedge \partial_{\phi_{1}}).
\end{equation}
Thus, the only relevant term on the left-hand side of~\eqref{eq:NCeq} is
\begin{equation}
\begin{aligned}
&\biggl[-\rho_{1} \Gamma_{\phi_{1}}(\Gamma_{\rho_{2}\phi_{2}} - \Gamma_{\rho_{3}\phi_{3}})\\
&\qquad\qquad+ \rho_{2}\Gamma_{\phi_{2}} (\Gamma_{\rho_{1}\phi_{1}}-\Gamma_{\rho_{3}\phi_{3}}) -\rho_{3}\Gamma_{\phi_{3}} (\Gamma_{\rho_{1}\rho_{1}}-\Gamma_{\rho_{2}\phi_{2}})\biggr] \epsilon=0\,,
\end{aligned}
\end{equation}
from which two independent conditions for the projectors are found:
\begin{equation}
 \pqty{\Gamma_{\rho_{1}\phi_{1}}-\Gamma_{\rho_{2}\phi_{2}}} \epsilon = \pqty{\Gamma_{\rho_{2}\phi_{}}-\Gamma_{\rho_{3}\phi_{3}}} \epsilon =0\,.
\end{equation} 
We see that this background preserves one quarter of the supersymmetries.
\end{example}

\section{Preserved Killing spinors for unimodular $r$-matrices of higher rank}
\label{sec:higher-rank}

In a first step, we show that with our formalism we can also determine the preserved Killing spinors for deformed backgrounds stemming from unimodular $r$-matrices of rank four, giving the explicit expressions for the projectors and the final resulting Killing spinors. The unimodular rank-4 $r$-matrices satisfying the \ac{cybe} were classified in Table~2 of~\cite{Borsato:2016ose}. We give here two examples from this list. Our results for the unbroken supersymmetries are consistent with the ones given in their Table~3.  

We also give an example of rank 6, but find that it does not preserve any supersymmetries.

Since the deformations appearing in this section act only on the \AdS{} part of the background, we do not need to consider the $S^{5}$-part of the Killing spinor and therefore do not give it explicitly in the final results.
For the \AdS{} part, we found it most convenient to write the \AdS{} background in Poincaré coordinates.
Thus, all the results are based on the following undeformed Killing spinor:
\begin{equation}
  \label{eq:undeformed-Killing}
\epsilon_{\mathrm{\AdS}_{5}} =  \left[\left(\sqrt{z} + \frac{1}{\sqrt{z}}x^{\alpha}\Gamma_{\alpha} \Gamma_{z}\right)\frac{1-i \gamma \Gamma_{z}}{2} + \frac{1}{\sqrt{z}}\frac{1+ i\gamma \Gamma_{z}}{2}\right](\epsilon_{0} + i \chi_{0})\,.
\end{equation}
\newpage
\begin{example}{$r= (M_{03}+M_{13}) \wedge (P_{0}+P_{1}) + P_{2}\wedge P_{3}$}
Let us consider the classical $r$-matrix\footnote{This example corresponds to $R_1$ in Table~2 of~\cite{Borsato:2016ose}.}
\begin{equation}
r= (M_{03}+M_{13}) \wedge (P_{0}+P_{1}) + P_{2}\wedge P_{3}\,.
\end{equation} 
The corresponding bi-vector $\Theta$ is given by 
\begin{equation}
\Theta = -2\lambda \left[(x^{0}-x^{1})(\partial_{0}+\partial_{1}) + \partial_{2}\right] \wedge \partial_{3}\,. 
\end{equation}
Note that one of the vectors is light-like so the projection formul

According to~\cite{Borsato:2016ose}, it preserves 8 supercharges. We now verify this result using our projector equation~\eqref{eq:projector}. The non-commutativity parameter $\Theta$ is divergence-free in the open string frame, consistently with r being unimodular:
\begin{equation}
\nabla_{\mu}\Theta^{\mu\nu} \hat{\Gamma}_{\nu}=0\,.
\end{equation}
The the l.h.s of the projector condition takes the form
\begin{equation}
\begin{aligned}
\left[ (x^{0}-x^{1})(1+\gamma \Gamma_{z01} \otimes (i\sigma_{2}))(\Gamma_{0} + \Gamma_{1}) + \Gamma_{2}(1 + \gamma \Gamma_{z} \otimes (i\sigma_{2}))\right]\epsilon=0\,.
\end{aligned}
\end{equation}
It is not hard to see that the above equation vanishes identically if the Killing spinor satisfies
\begin{equation}
(1 + \gamma \Gamma_{z}\otimes(i\sigma_{2}))\epsilon=0\quad \mathrm{and}\quad (\Gamma_{0} + \Gamma_{1})\epsilon=0\,,
\end{equation}
leading us to the conclusion that indeed only $\frac{1}{4}$-\ac{susy} is preserved.
\end{example}
Explicit calculations lead to 
\begin{equation}
\frac{1}{2}\Theta^{mn}\Gamma_{mn} = \frac{\lambda}{z^{2}}\left[ (x^{0}-x^{1})(\Gamma_{0} + \Gamma_{1})+ \Gamma_{2}\right]\Gamma_{3}
\end{equation}
and
\begin{equation}
\omega(\Theta) = \frac{z^{2}}{\lambda}\arctan \left(\frac{\lambda}{z^{2}}\right)\,.
\end{equation}
The resulting preserved Killing spinors of the deformed background are therefore given by
\begin{equation}
  \begin{aligned}
    \epsilon_{+}^{\text{(fin)}} &= \exp\left[\arctan\left(\frac{\lambda}{z^{2}}\right) \left((x^{0}-x^{1})(\Gamma_{0} + \Gamma_{1})+\Gamma_{2}\right)\Gamma_{3}\right]\epsilon_{+}\\
    \epsilon_{-}^{\text{(fin)}} &= \epsilon_{-}\,,
  \end{aligned}
\end{equation}
where 
\begin{equation}
\epsilon_{+} + i\epsilon_{-} = \frac{1}{\sqrt{z}}\frac{1+ i\gamma \Gamma_{z}}{2}\frac{1+ \Gamma_{01}}{2}\epsilon_{0}\,.
\end{equation}

\begin{example}{$r= P_{1} \wedge (M_{02} - M_{23}) + a (P_{0} + P_{3}) \wedge (D-M_{03})$}
Also this case is listed in~\cite{Borsato:2016ose}\footnote{It corresponds to $R_5$ in Table~2 of~\cite{Borsato:2016ose}.}. We can read off the non-commutativity bi-vector from the classical $r$-matrix:
\begin{multline}
\Theta = \lambda \biggl[ (-x^{0}+x^{3})\partial_{1} \wedge \partial_{2} + x^{2} \partial_{1} \wedge (\partial_{0} + \partial_{3}) \\
 + a ( 2 (x^{0}- x^{3}) \partial_{0} \wedge \partial_{3} + (x^{1}\partial_{1} + x^{2}\partial_{2} + z \partial_{z}) \wedge (\partial_{0} + \partial_{3}) )\biggr] .
\end{multline}
Again, the non-commutativity parameter satisfies $\nabla_{\mu}\Theta^{\mu\nu} =0.$ Therefore, the projector equation can be written as
\begin{equation}
\begin{aligned}
&\biggl[\frac{z}{2}\Gamma_{12}(\Gamma_{0}+\Gamma_{3}) + (x^{0}-x^{3})\Gamma_{12z}(1 + \gamma \Gamma_{z} \otimes (i\sigma_{2})) -x^{2}(1 + \gamma \Gamma_{z03} \otimes (i\sigma_{2})) (\Gamma_{0}+\Gamma_{3})\\
&\qquad + a (-z \Gamma_{12}(\Gamma_{0} + \Gamma_{3}))\otimes (i\sigma_{2})-2 (x^{0}-x^{3})\Gamma_{03z}(1 + \gamma \Gamma_{z} \otimes (i\sigma_{2}))\\
&\qquad \qquad +(x^{1} + x^{2}) \Gamma_{2z} (1 + \gamma \Gamma_{z03} \otimes (i\sigma_{2}))(\Gamma_{0} + \Gamma_{3})\biggr]\epsilon =0\,.
\end{aligned}
\end{equation}
We can easily find two projection conditions to find nontrivial Killing spinors:
\begin{align}
(\Gamma_{0} + \Gamma_{3}) \epsilon &=0 \,, & (1 + \gamma \Gamma_{z} \otimes (i\sigma_{2})) \epsilon &= 0\,.
\end{align}
We see that we are left with $\frac{1}{4}$-supersymmetry because we have two projectors,
\begin{align}
  \Pi_{1} &= \frac{1+i\gamma\Gamma_{z}}{2} \,, & \Pi_{2} &= \frac{1 +\Gamma_{03}}{2}\,.
\end{align}
Furthermore,
\begin{multline}
  \frac{1}{2}\Theta^{mn}\Gamma_{mn} = \frac{\lambda}{z^{2}}\biggl[ \Gamma_{1}((x^{3}-x^{0})\Gamma_{2}- x^{2}(\Gamma_{0} +\Gamma_{3})) \\+ a \pqty{2 (x^{0}-x^{3})\Gamma_{03} +  \pqty{x^{1} \Gamma_{1} + x^{2} \Gamma_{2}+z \Gamma_{z}} \pqty{\Gamma_{0} + \Gamma_{3}}}\biggr]
\end{multline} 
and
\begin{equation}
\frac{1}{2}\Theta^{mn}\Theta_{mn} = \frac{\lambda^{2}}{z^{4}}(1-4a^{2})(x^{0}-x^{3})^{2}
\end{equation} 
\end{example}
give rise to the following preserved Killing spinors:
\begin{equation}
\begin{aligned}
  \epsilon_{+}^{\text{(fin)}} &= \exp[\frac{-\arctan(\lambda\sqrt{1-4a^{2}}(x^{0}-x^{3})/z^{2})}{\sqrt{1-4a^{2}}(x^{0}-x^{3})} \frac{z^2}{2 \lambda} \Theta^{mn} \Gamma_{mn} %
  ] \epsilon_{+}\\
\epsilon_{-}^{\text{(fin)}} &= \epsilon_{-}\,,
\end{aligned}
\end{equation}
where 
\begin{equation}
\epsilon_{+} + i \epsilon_{-} = \frac{1}{\sqrt{z}} \frac{1+i\gamma\Gamma_{z}}{2}\frac{1 +\Gamma_{03}}{2} \epsilon_{0}\,.
\end{equation}

\begin{example}{$ r = P_{0} \wedge P_{1} + P_{2} \wedge P_{3} + M_{01} \wedge M_{23}$}
Let us next study a case of a rank six. Also this example comes from~\cite{Borsato:2016ose}. Intuitively, judging from the last term, which would work as a TsT without mixing angles alone, the deformation with this classical $r$-matrix is expected to break all supersymmetries.

Indeed, the matrix in the  projector equation is given by
\begin{multline}
-\frac{4}{z}\Gamma_{23z}+(x^{0} \Gamma_{123} + x^{1} \Gamma_{023} + x^{2} \Gamma_{013} -x^{3} \Gamma_{012}) + \\
\frac{2}{z}\Gamma_{z} \biggl[x^{0}x^{3} (\Gamma_{03}-\Gamma_{12}) + x^{0}x^{2} (\Gamma_{02} + \Gamma_{13} ) + \\
x^{1} x^{2} ( \Gamma_{03} + \Gamma_{12}) + x^{1} x^{3} (-\Gamma_{02} + \Gamma_{13})\biggr]
\end{multline}
and it has a non-zero determinant.
Thus, supersymmetry is completely broken.
It would be interesting to find a supersymmetric case of rank six.
\end{example}

\section{Killing spinor formula in generalized SUGRA/for non-unimodular $r$-matrices}
\label{sec:generalized-SUGRA}

We now turn to backgrounds related to \acp{ybd} with non-unimodular $r$-matrices. These backgrounds obey the generalized supergravity equations.

\begin{example}{$r= (P_{0}-P_{1}) \wedge (D-L_{01})$}
This example was first discussed in~\cite{Orlando:2016qqu}. 
The background given below satisfies the generalized supergravity equations of motion with a non-trivial Killing vector field $I^\mu$ given by the divergence formula~\cite{arXiv:1702.02861}:
\begin{equation*}
\begin{aligned}
\dd s^{2} &= \frac{-(\dd x^{0})^{2}+(\dd x^{1})^{2}+(\dd x^{2})^{2}+(\dd x^{3})^{2}+\dd z^{2}}{z^{2}} - \lambda^{2}\frac{(x^{2})^{2}+(x^{3})^{2}+z^{2}}{2z^{6}}(\dd x^{0}+\dd x^{1})^{2}\\
&\qquad\quad+\dd\theta_{2}^{2}+ \sin^{2}\theta_{2}\dd\theta_{1}^{2}+\cos^{2}\theta_{1}\sin^{2}\theta_{2}\dd\phi_{1}^{2} + \sin^{2}\theta_{1}\sin^{2}\theta_{2}\dd\phi_{2}^{2} + \cos^{2}\theta_{2}\dd\phi_{3}^{2}\,,\\
B_{2} &= \frac{\lambda}{2\sqrt{2} z^{4}}\dd(x^{0}+x^{1})\wedge \dd\left[(x^{2})^{2}+(x^{3})^{2}+z^{2})\right]\,,
\end{aligned}
\end{equation*}
\begin{equation}
  \begin{aligned}
    F_{3} &= \frac{2\sqrt{2}\lambda}{z^{5}}\dd(x^{0}+x^{1}) \wedge (x^{2} \dd x^{3} \wedge \dd z + x^{3} \dd z \wedge \dd x^{2} + z \dd x^{2}  \wedge \dd x^{3})\,,\\
F_{5} &= 4\left[\omega_{AdS_{5}}+\omega_{S^{5}} \right]\,,\\
I^{m}\partial_{m} &= \sqrt{2}\lambda(\partial_{0} -\partial_{1})\,,\\
\omega_{AdS_{5}} & = -\frac{\dd x^{0}\wedge \dd x^{1}\wedge \dd x^{2}\wedge \dd x^{3}\wedge \dd z}{z^{5}}\,,\\
 \omega_{S^{5}} &= -\cos\theta_{1}\sin\theta_{1}\cos\theta_{2}\sin^{3}\theta_{2} \dd\theta_{1}\wedge \dd\theta_{2}\wedge \dd\phi_{1} \wedge \dd\phi_{2} \wedge \dd\phi_{3}\,.
\end{aligned}
\end{equation}
To obtain the antisymmetric bi-vector, we can either formally apply the \ac{sw} map or construct the bi-Killing structure using the classical $r$-matrix:  
\begin{equation}
\Theta = \frac{-\lambda}{\sqrt{2}} (\partial_{0} - \partial_{1}) \wedge (x^{2}\partial_{2} + x^{3}\partial_{3} + z \partial_{z})\,.
\end{equation}
Let us look at the number of unbroken supersymmetries. The projector equation~\eqref{eq:projector} can be written as
\begin{equation}
(x^{2}\Gamma_{2} + x^{3} \Gamma_{3} + z \Gamma_{z} )\Gamma_{z} (1- \gamma \Gamma_{z01} \otimes (i\sigma_{2}))(\Gamma_{0}- \Gamma_{1})\epsilon = 0\,,
\end{equation} 
where $\gamma = \Gamma_{56789}$. We see that the Killing spinor is required to satisfy
\begin{equation}
\label{eq:proj1}
(\Gamma_{0} - \Gamma_{1})\epsilon = 0\,. 
\end{equation}

In this configuration the bi-vector is written as the wedge product of a light-like Killing spinor and a space-like one.
As we have already observed in Section~\ref{sec:KosmannTheta}, this means that Eq.~(\ref{eq:NCeq}) is just a necessary condition and we need to check explicitly if the \ac{kl} derivative in the directions of the Killing vectors vanishes.
The result is that all supersymmetries are broken since in the frame that we have chosen there is no Killing spinor of the undeformed \(\AdS{5} \times S^5\) background that is independent of \(z\).

\end{example}

\begin{example}{$r=P_{1}\wedge D$}
The full background data of this example was also given in~\cite{Orlando:2016qqu}, and satisfies again the generalized supergravity equations of motion. The non-commutativity bi-vector in this case reads
 \begin{equation}
 \Theta = \lambda \partial_{1} \wedge (-x^{0} \partial_{0} -x^{2} \partial_{2} - x^{3}\partial_{3} - z \partial_{z})\,.
 \end{equation}
We write \eqref{eq:NCeq} in Poincaré-coordinates:  
\begin{equation}
\biggl[\frac{4}{z}\Gamma_{1}+ \frac{8}{z}\Gamma_{023}\otimes (i\sigma_{2}) + \frac{8}{z^{2}} \biggl\{(-x^{0}\Gamma_{0} - x^{2}\Gamma_{2} - x^{3} \Gamma_{3})\Gamma_{1z} (1-\Gamma_{0123}\otimes(i\sigma_{2}))\biggr\}\biggr] \epsilon =0\,.
\end{equation}
In this case, one can explicitly verify that the determinant of the matrix acting on $\epsilon$ does not vanish and the only solution to the above equation is $\epsilon = 0$. No supercharges are preserved by the deformed background.
\end{example}

\section{Killing spinor formula for non-TsT $r$-matrices}\label{sec:nonTsT}

Despite our derivation relying on TsT transformations, we conjecture that our results are valid also for more general cases. As a first check, we apply our formula to three examples not corresponding to TsT transformations given in~\cite{Borsato:2016ose}\footnote{They correspond to the cases $R_{15}$, $R_{16}$ and $R_{17}$ in Table~2.}. Again, our results for the amount of unbroken supersymmetries are consistent with the ones given in Table~3 of~\cite{Borsato:2016ose}.  

\begin{example}{$r= P_{1} \wedge P_{3} + (P_{0}+P_{1}) \wedge (M_{03}+M_{13}) $ }
The non-commutativity bi-vector of this example can be explicitly written as
\begin{equation}
\Theta = -2\lambda \partial_{1} \wedge \partial_{3} + 2\lambda(x^{0}-x^{1}) (\partial_{0} + \partial_{1}) \wedge \partial_{3} + \mathcal{O}(\lambda^{2}).
\end{equation}
Note that this bi-vector $\Theta_{mn}$ is divergence-free. 
Factorizing the $\Gamma$-matrices, we obtain the projector equation
\begin{equation}
\frac{16\lambda}{z^{2}}\Gamma_{3z}\left[\Gamma_{1}(1+\gamma \Gamma_{z} \otimes (i\sigma_{2}))-(x^{0}-x^{1})(1+\gamma \Gamma_{z}\Gamma_{01} \otimes (i\sigma_{2})) (\Gamma_{0} + \Gamma_{1})\right]\epsilon = 0.
\end{equation}
It follows from 
\begin{equation}
(1+\gamma \Gamma_{z}\otimes (i\sigma_{2}))\epsilon=0\,,\qquad (1-\Gamma_{01})\epsilon=0\,.
\end{equation}
This non-zero solution is restricted by the two projectors
\begin{equation}
\Pi_{1} = \frac{1}{2}(1+i\gamma \Gamma_{z})\,, \qquad \Pi_{2} =  \frac{1}{2}(1+\Gamma_{01})\,,
\end{equation}
preserving 8 supercharges. 

Since for this deformation, we cannot rely on the proof given in Section~\ref{sec:kosmann-lie}, we will explicitly show that the Killing spinors given by our construction are preserved. For simplicity, we focus on the linear order of the deformation, so that the metric of AdS${}_{5} \times S^{5}$ remains undeformed. Therefore the only non-trivial contribution of the deformation arises from the $B_{2}$ field and $F_{3}$ flux whereas the $F_{5}$ flux is undeformed:
\begin{equation}
\begin{aligned}
B_{2} &= \frac{2\lambda}{z^{4}}\left[(x^{0}-x^{1}) \dd(x^{0}-x^{1}) + \dd x^{1} \right] \wedge \dd x^{3} + \mathcal{O}(\lambda^{2})\,,\\
F_{3} &= \frac{8\lambda}{z^{5}}\left[(x^{0}-x^{1}) \dd (x^{0}-x^{1}) - \dd x^{0}\right] \wedge \dd x^{2} \wedge \dd z + \mathcal{O}(\lambda^{2}).\\
\end{aligned}
\end{equation}
Also, we find 
\begin{equation}
\frac{1}{2}\Theta_{mn}\Theta^{mn} = \frac{4\lambda^{2}}{z^{4}}(1- 2(x^{0} - x^{1}))\,,
\end{equation}
\begin{equation}
\frac{1}{2}\Theta_{mn}\Gamma^{mn} = \frac{2\lambda}{z^{2}}\left[ (x^{0}-x^{1})(\Gamma_{0}+\Gamma_{1})- \Gamma_{1}\right]\Gamma_{3}\,,
\end{equation}
and
\begin{equation}
\omega(\Theta) = 1+ \mathcal{O}(\lambda^{2}).
\end{equation}
Thus, at leading order in $\lambda$, the final Killing spinor is written as
\begin{align}
  \epsilon_{+}^{\text{(fin)}} &= \left[1 + \frac{2\lambda}{z^{2}}( (x^{0} - x^{1})(\Gamma_{0}+\Gamma_{1}) - \Gamma_{1})\Gamma_{3}\right] \epsilon_{+} \,, \\ \epsilon_{-}^{\text{(fin)}} &= \epsilon_{-}\,,
\end{align}
where 
\begin{equation}
\epsilon_{+} + i \epsilon_{-} = \frac{1}{\sqrt{z}}\Pi_{1}\Pi_{2}\epsilon_{0}.
\end{equation}

Let us first consider the dilatino variation. Using the background data, we find
\begin{equation}
\begin{aligned}
\slashed{H}_{3} &= \frac{48\lambda}{z^{2}}\Gamma_{3z} \left[ (x^{0}-x^{1})(\Gamma_{0} + \Gamma_{1}) - \Gamma_{1}  \right] = -3\nabla_{m}\Theta^{np}\Gamma^{m}{}_{np} \,,\\
\Gamma_{m}\mathcal{S}\Gamma^{m} &=-\frac{32\lambda}{z^{2}} \Gamma_{2z}\left[ (x^{0}-x^{1})(\Gamma_{0}+\Gamma_{1}) - \Gamma_{0}\right] \otimes \sigma_{1} = 2 \Theta^{mn}\Gamma_{m}\mathcal{S}_{\mathrm{o}}\Gamma_{n} \otimes \sigma_{3}\,.
\end{aligned}
\end{equation}
Therefore, the dilatino variation takes the form 
\begin{equation}
\begin{aligned}
\delta \lambda &= -\frac{1}{24}\slashed{H}_{3} \otimes \sigma_{3} \tilde{\epsilon} + \frac{1}{16}\Gamma_{m}\mathcal{S}\Gamma^{m}\tilde{\epsilon},\\ 
&= \frac{1}{8} \left[ \Theta^{mn} \Gamma_{m}\mathcal{S}\Gamma_{n} + \nabla_{m}\Theta^{np}\Gamma^{m}{}_{np} \right] \otimes \sigma_{3}\tilde{\epsilon} = 0\,.
\end{aligned}
\end{equation}
As expected, we recover the projector formula. Explicit computations show that the dilatino variation vanishes. %

As for the gravitino variation, let us consider the component $\delta \Psi_{0}$.
Using the background data, we compute at linear order in $\lambda$:
\begin{equation}
\begin{aligned}
\nabla_{0} &= \partial_{0} - \frac{1}{2z}\Gamma_{0z}, \\
\frac{1}{8}H_{0mn}\Gamma^{mn} &= -\frac{2\lambda}{z^{3}} (x^{0}-x^{1})\gamma\Gamma_{012},\\
\frac{1}{8}\mathcal{S}\Gamma_{0} &= -\frac{\lambda}{z^{3}} \Gamma_{2z} \left[ (x^{0}-x^{1}) (1+ \Gamma_{01}) - 1 \right] \otimes \sigma_{1} - \frac{1}{2z} \gamma \Gamma_{0} \otimes (i\sigma_{2})\,,
\end{aligned}
\end{equation} 
where again $\gamma = \Gamma_{\theta_{1}\theta_{2}\phi_{1}\phi_{2}\phi_{3}} = \Gamma_{56789}.$ Thus, using the Killing spinor given in~(\ref{eq:undeformed-Killing}), we find
\begin{equation}
\begin{aligned}
\delta \Psi_{0 +} &= \nabla_{0}\epsilon_{+}^{\text{(fin)}} - \frac{1}{8}H_{0mn}\Gamma^{mn}\epsilon_{+}^{\text{(fin)}} + \frac{1}{8}\left[-\frac{1}{6}\slashed{\mathcal{F}}_{3}\Gamma_{0} \epsilon_{-}^{\text{(fin)}}  - \frac{1}{240} \slashed{\mathcal{F}}_{5}\Gamma_{0} \epsilon_{-}^{\text{(fin)}}\right]\\
&= \frac{2\lambda}{z^{3}} (x^{0}-x^{1})\gamma\Gamma_{012} \epsilon_{+}^{\text{(fin)}} -\frac{\lambda}{z^{3}} \Gamma_{2z} \left[ (x^{0}-x^{1}) (1+ \Gamma_{01}) + 1\right]\epsilon_{-}^{\text{(fin)}} = 0.
\end{aligned}
\end{equation}
We can see analogously that also the other components vanish.

\end{example}

\begin{example}{$r=(P_{1}+a(P_{0}+P_{3})) \wedge P_{3} + (P_{2}+M_{01}-M_{23})\wedge (P_{0}+P_{3})$}
For this example, the non-commutativity can be simply read off as
\begin{equation}
\Theta = \lambda (\partial_{1} + a \partial_{0}) \wedge \partial_{3} + \lambda  (x^{0} - x^{3}) (\partial_{0} + \partial_{3}) \wedge \partial_{1} + \lambda \partial_{2} \wedge (\partial_{0} + \partial_{3})\,.
\end{equation}
We can extract the projectors by writing down \eqref{eq:NCeq} explicitly:
\begin{equation}
\begin{aligned}
&\biggl[(\Gamma_{02} - \Gamma_{13} - \Gamma_{23} - a \Gamma_{03})(1 + \gamma \Gamma_{z} \otimes (i\sigma_{2}))\\
&\qquad+(x^{0}-x^{3})(1- \gamma \Gamma_{z} \Gamma_{03} \otimes (i\sigma_{2}))(\Gamma_{0} + \Gamma_{3})\biggr]\epsilon =0\,.
\end{aligned}
\end{equation}
To get a non-zero solution, we need to impose the two projections:
\begin{equation}
(\Gamma_{0} + \Gamma_{3})\epsilon = (1 + \gamma \Gamma_{z} \otimes (i\sigma_{2}) )\epsilon = 0 \,,
\end{equation} 
which implies that 8 supercharges are preserved. 
Since
\begin{equation}
\frac{1}{2}\Theta^{mn}\Gamma_{mn} = \frac{\lambda}{z^{2}}\biggl[\Gamma_{13} + a \Gamma_{03} + \biggl\{(x^{3} - x^{0})\Gamma_{1} + \Gamma_{2}\biggr\}(\Gamma_{0} + \Gamma_{1})\biggr] \,,
\end{equation}
the resulting Killing spinor is given by
\begin{equation}
\begin{aligned}
\epsilon_{+}^{\text{(fin)}} &= \exp\biggl[ \frac{\arctan\sqrt{|\Delta|}}{\sqrt{|\Delta|}} \frac{\lambda}{z^{2}}\biggl(\Gamma_{13} + a \Gamma_{03} + \biggl\{(x^{3} - x^{0})\Gamma_{1} + \Gamma_{2}\biggr\}(\Gamma_{0} + \Gamma_{1})\biggr)\biggr]\epsilon_{+}\,,\\
\epsilon_{-}^{\text{(fin)}} &=\epsilon_{-}\,,
\end{aligned}
\end{equation}
where
\begin{equation}
\Delta = \frac{1}{2}\Theta_{mn}\Theta^{mn} = \frac{\lambda^{2}}{z^{4}} (1-a^{2} - 2 x^{0} + 2 x^{3})\,,
\end{equation}
and
\begin{equation}
\epsilon_{+} + i \epsilon_{-} = \frac{1}{\sqrt{z}}\frac{1}{2}(1+i\gamma \Gamma_{z}) \frac{1}{2}(1+\Gamma_{03})\epsilon_{0}\,.
\end{equation}
\end{example}

\begin{example}{$r=(P_{1}+a(P_{0}+P_{3})) \wedge (P_{1}+P_{3}+M_{02}-M_{23}) + (P_{0}+P_{3})\wedge (P_{2}+M_{01}-M_{13})$}
By reading off the non-commutativity from the above $r$-matrix as
\begin{equation}
\begin{aligned}
\Theta &= \lambda (\partial_{1} + a (\partial_{0}+\partial_{3})) \wedge (\partial_{1}+\partial_{3} - x^{0}\partial_{2} - x^{2}\partial_{0}-x^{2}\partial_{3} + x^{3}\partial_{2})\\
 &\qquad+\lambda (\partial_{0}+\partial_{3}) \wedge (\partial_{2} - x^{0}\partial_{1} - x^{1}\partial_{0} - x^{1}\partial_{3} + x^{3}\partial_{1})\,,
 \end{aligned}
\end{equation}
one can factorize the matrices in the projector equation as follows:
\begin{multline}
\biggl[2z \Gamma_{12}(\Gamma_{0} + \Gamma_{3}) - \Gamma_{z} (\Gamma_{02} - \Gamma_{23} + \Gamma_{13})(1 + \gamma \Gamma_{z} \otimes (i\sigma_{2}))\\
+(x^{0}-x^{3})\Gamma_{z} (\Gamma_{01} + \Gamma_{12} - \Gamma_{13})(1+\gamma \Gamma_{z} \otimes(i\sigma_{2}))\\
x^{2}\Gamma_{1z}(1+\gamma \Gamma_{z} \Gamma_{03} \otimes (i\sigma_{2}))(\Gamma_{0}+\Gamma_{3})\\
+a\biggl( -\Gamma_{z} (\Gamma_{01} + \Gamma_{03} - \Gamma_{13})(1+\gamma \Gamma_{z} \otimes (i\sigma_{2})) \\
+(x^{0}-x^{3}) \Gamma_{2z} (1+\gamma\Gamma_{z} \Gamma_{03} \otimes (i\sigma_{2}))(\Gamma_{0} + \Gamma_{3})\biggr)\biggr]\epsilon=0\,.
\end{multline}
We can find two independent projections, and obtain
\begin{equation}
(\Gamma_{0} + \Gamma_{3}) \epsilon = (1 + \gamma \Gamma_{z} \otimes (i\sigma_{2}))\epsilon=0\,,
\end{equation}
which implies 8 supercharges.
Using the undeformed Killing spinor in Poincare coordinates and 
\begin{equation}
\begin{aligned}
\frac{1}{2}\Theta^{mn}\Gamma_{mn} &= \frac{\lambda}{z^{2}}\biggl[(\Gamma_{02}+\Gamma_{13}-\Gamma_{23})-(x^{0}-x^{3})\Gamma_{12}+(x^{0}-x^{2}-x^{3})\Gamma_{1}(\Gamma_{0}+\Gamma_{3})\\
&\qquad+ a \biggl(\Gamma_{01} + \Gamma_{03}-\Gamma_{13} + (x^{0}-x^{3})\Gamma_{2} (\Gamma_{0} + \Gamma_{3})\biggr)\biggr]\,,
\end{aligned}
\end{equation}
we can directly construct the Killing spinor
\begin{equation}
\begin{aligned}
  \epsilon_{+}^{\text{(fin)}} &= \exp[\frac{\arctan\sqrt{|\Delta|}}{2\sqrt{|\Delta|}}\frac{\lambda}{z^{2}} \Theta^{mn}\Gamma_{mn}%
  ] \epsilon_{+},\\
\epsilon_{-}^{\text{(fin)}} &=\epsilon_{-}\,,
\end{aligned}
\end{equation}
where 
\begin{equation}
\epsilon_{+} + i \epsilon_{-} = (1+i\gamma \Gamma_{z})(1+\Gamma_{03})\epsilon_{0}\,,
\end{equation}
and 
\begin{equation}
\Delta = \frac{1}{2} \Theta^{mn}\Theta_{mn} = \frac{\lambda^{2}}{z^{4}}\biggl[2-(1+a)^{2} + 2(x^{0}-x^{2}-x^{3}) + (x^{0}-x^{3})^{2}\biggr].
\end{equation}
\end{example}

\vspace{-3em}
\section{Discussion and Conclusions}
\label{sec:conclusions}
In the study of supersymmetric gauge theories constructed from deformed string theory backgrounds, the explicit expressions of the preserved Killing spinors are an important technical ingredient. %

In this paper, we have given a frame-independent formulation for the preserved Killing spinors of a deformed ten-dimensional background which is related a known background (such as  $\AdS{5}\times S^5$ or flat space) via sequences of consecutive TsT-transformations or generalized T-dualities.
We explicitly check our formula on a variety of backgrounds corresponding to integrable deformations that can be encoded in terms of an $r$-matrix.
These examples include cases of higher rank $r$-matrices and non-unimodular $r$-matrices leading to backgrounds in generalized supergravity.
Even though our construction uses the properties of the TsT transformation, the final result is expressed solely in terms of the anti-symmetric bi-vector $\Theta$, obtained by the generalized Seiberg--Witten map.
This result motivates us to conjecture that our formula is applicable beyond the TsT case. We checked this conjecture explicitly on a number of non-TsT examples.

\medskip

There is a number of open questions that suggest themselves for future work:
\begin{itemize}
\item It would be highly desirable to confirm our projector formula in
  a different framework, independent of Abelian T-duality. A natural
  setting could be the \ac{susy} variations in \(\beta\)-supergravity
  recently discussed in~\cite{Bakhmatov:2018bvp}.
\item It would also be interesting to delineate exactly the range of validity of our expressions.
Deformations not satisfying the Jacobi identity, for example, have not yet been considered in this context and will probably require an extra term in the projection equation.
On general grounds we expect this modification to be related to the non-geometric $R$-flux. %
\item Finally, it might be possible to obtain a clearer geometric interpretation of our construction in terms of a twist of the derivative structure by the non-geometric fluxes~\cite{Andriot:2014qla}.%
\end{itemize}
\subsection*{Acknowledgments}

It is our pleasure to thank E.~Ó~Colgáin and J.~Sakamoto for fruitful discussions.
D.O. and S.R. gratefully acknowledge
support from the Simons Center for Geometry and Physics, Stony Brook University at which some of the research for this paper was performed.  
Y.S. would like to thank École normale supérieure Paris, KU Leuven, Université libre de Bruxelles, Ludwig-Maximilian University of Munich, and Humboldt University for hospitality.
The work of S.R. and Y.S. is supported by the Swiss National Science Foundation (\textsc{snf}) under grant number PP00P2\_157571/1.
D.O.~acknowledges partial support by the NCCR 51NF40-141869 ``The Mathematics of Physics'' (SwissMAP). 
The work of K.Y. is supported by the Supporting Program for Interaction based Initiative Team Studies 
(SPIRITS) from Kyoto University and by a JSPS Grant-in-Aid for Scientific Research (B) No.\,18H01214.   
This work is also supported in part by the JSPS Japan-Russia Research Cooperative Program.  
\appendix
\section{$\Gamma$-matrices}
\label{app:Gamma}
We parametrize the Gamma matrices based on the conventions in~\cite{Lu:1998nu}. The five-dimensional Gamma matrices are defined as
\begin{equation}
\label{eq:Gamma-matrix}
\begin{aligned}
\Gamma_{1} &= \sigma_{1} \otimes \mathbf{1} \otimes \mathbf{1} \otimes\cdot\cdot\cdot \otimes\mathbf{1}\,,\quad &\Gamma_{2} &= \sigma_{2} \otimes \mathbf{1} \otimes \mathbf{1} \otimes\cdot\cdot\cdot \otimes \mathbf{1}\,,\\
\Gamma_{3} &= \sigma_{3} \otimes \sigma_{1} \otimes \mathbf{1} \otimes \cdot\cdot\cdot \otimes\mathbf{1}\,,\quad &\Gamma_{4} &= \sigma_{3} \otimes \sigma_{2} \otimes \mathbf{1} \otimes \cdot\cdot\cdot \otimes \mathbf{1}\,,\\
\cdot\cdot\cdot\\
\Gamma_{2n-1} &= \sigma_{3} \otimes \sigma_{3} \otimes \cdot\cdot\cdot \otimes \sigma_{3} \otimes \sigma_{1}\,,\quad & \Gamma_{2n} &= \sigma_{3} \otimes \sigma_{3} \otimes \cdot\cdot\cdot \otimes \sigma_{3} \otimes \sigma_{2},
\end{aligned}
\end{equation}
where $\sigma_i~(i=1,2,3)$ are the standard Pauli matrices. 
The imaginary $i$ appears in front when considering a metric with Lorentzian signature.
It is convenient to decompose the ten-dimensional Dirac matrices into lowe$r$-dimensional ones and to use the latter in the calculations. In the case of $\AdS{5} \times S^{5}$\,, we decompose the ten-dimensional Gamma matrices, denoted by $\hat{\Gamma}_{A}$, as follows:
\begin{equation}
\label{eq:gamma-decomposition}
\begin{aligned}
\AdS{5}:& \quad \ \hat{\Gamma}_{\mu} =\,  \sigma_{1} \otimes \Gamma_{\mu} \otimes \mathbf{1}_{4\times4}\,,\\
S^{5}: & \quad \ \hat{\Gamma}_{m} =\, \sigma_{2} \otimes \mathbf{1}_{4\times4} \otimes \Gamma_{m}\,,
\end{aligned}
\end{equation}
where {the $\sigma_{1,2}$ are needed} to satisfy the Clifford algebra in ten dimensions. 
Using this parametrization, the ten-dimensional chirality matrix is explicitly written as
\begin{equation}
\begin{aligned}
\label{eq:chirality-matrix}
\hat{\Gamma}_{(10)} &= \hat{\Gamma}_{0} \cdot\cdot\cdot \hat{\Gamma}_{9}\\
&=\sigma_{3} \otimes \mathbf{1}_{4\times4} \otimes \mathbf{1}_{4\times4}\,,
\end{aligned}
\end{equation}
which automatically gives a positive chirality for the Killing spinor of the form
\begin{equation}
\label{eq:spinor-decomposition}
\epsilon =
\begin{pmatrix}
1\\0
\end{pmatrix}
\otimes \epsilon_{\AdS{5}}
\otimes \epsilon_{S^{5}} .
\end{equation}
We furthermore give another set of ten-dimensional Gamma matrices with only real components:
\begin{equation}
\begin{aligned}
\Gamma_{0} &= i\sigma_{2} \otimes \mathbf{1}_{4\times4} \otimes \mathbf{1}_{4\times4}\,,\quad &\Gamma_{1} &= \sigma_{1} \otimes \sigma_{2} \otimes \sigma_{2} \otimes \sigma_{2} \otimes \sigma_{2}\,, \\
\Gamma_{2} &= \sigma_{1} \otimes \sigma_{2} \otimes \mathbf{1}_{2\times2} \otimes \sigma_{1} \otimes \sigma_{2}\,,\quad  &\Gamma_{3} &=\sigma_{1} \otimes \sigma_{2} \otimes \mathbf{1}_{2\times 2} \otimes \sigma_{3} \otimes \sigma_{2}\,,\\
\Gamma_{4} &=\sigma_{1} \otimes \sigma_{2} \otimes \sigma_{1} \otimes \sigma_{2} \otimes \mathbf{1}_{2\times2}\,,\quad &\Gamma_{5} &= \sigma_{1} \otimes \sigma_{2} \otimes \sigma_{3} \otimes \sigma_{2} \otimes \mathbf{1}_{2\times 2}\,,\\
\Gamma_{6} &= \sigma_{1} \otimes \sigma_{2} \otimes \sigma_{2} \otimes \mathbf{1}_{2\times2} \otimes \sigma_{1}\,,\quad &\Gamma_{7} &= \sigma_{1} \otimes \sigma_{2} \otimes \sigma_{2}\otimes \mathbf{1}_{2\times2} \otimes \sigma_{3}\,,\\
\Gamma_{8} &=\sigma_{1}\otimes \sigma_{1} \otimes \mathbf{1}_{2\times2} \times \mathbf{1}_{2\times2} \otimes \mathbf{1}_{2\times2}\,,\quad &\Gamma_{9} &=\sigma_{1}\otimes \sigma_{3} \otimes \mathbf{1}_{2\times2} \otimes \mathbf{1}_{2\times2}\otimes \mathbf{1}_{2\times2}\,.
\end{aligned}
\end{equation}
With these Gamma matrices, the chirality matrix takes the same form as \eqref{eq:chirality-matrix}.
\section{Killing spinors for undeformed backgrounds}
This appendix deals with the solutions of the gravitino equation for the $\AdS{5}\times S^{5}$ background, 
\begin{equation}
D_{m} \epsilon + \frac{i}{8\cdot 2\cdot 5!}\slashed{F}_{5}\hat{\Gamma}_{m}\epsilon=0.
\end{equation}
Using the decomposition of the Gamma matrices as well as the Killing spinors, we decompose the Killing equation into two sectors:
\begin{equation}
\begin{aligned}
\AdS{5}:& \quad \ D_{\mu}\epsilon_{\AdS{5}} =\,  \frac{1}{2}\Gamma_{\mu} \epsilon_{\AdS{5}}\,,\\
S^{5}: & \quad \ D_{j}\epsilon_{S^{5}} =\, \frac{i}{2}\Gamma_{j}\epsilon_{S^{5}}\,,
\end{aligned}
\end{equation}
where both $\Gamma_{\mu}$ and $\Gamma_{j}$ are five-dimensional representations.
In particular, depending on which classical $r$-matrices we choose, it is convenient to introduce different coordinates for $\AdS{5}$.

\subsection{$\AdS{5}$ in Poincar\'e coordinates}
Let us first express the metric as
\begin{equation}
ds^{2}_{\AdS{5}} = \frac{\eta_{\alpha\beta}dx^{\alpha}dx^{\beta}  + (dz)^{2}}{z^{2}}\,,
\end{equation}
where $\eta_{\alpha\beta} = \mathrm{diag}(-,+,+,+)$. A natural choice for the vielbein one-form is given by
\begin{equation}
\mathbf{e}^{\alpha} = \frac{dx^{\alpha}}{z}\,, \qquad \mathbf{e}^{z} = \frac{dz}{z}\,.
\end{equation}
Then one finds the spin connections 
\begin{equation}
\omega_{\alpha}^{z \alpha} = \frac{1}{z}\,,\qquad \omega_{z}^{z \alpha} = 0\,,
\end{equation}
and the gravitino equation reads
\begin{equation}
\begin{aligned}
\partial_{\alpha} \epsilon &= \frac{1}{2z}\Gamma_{\alpha}(1+\Gamma_{z})\epsilon\,,\\
\partial_{z}\epsilon &= \frac{1}{2z} \Gamma_{z}\epsilon\,.
\end{aligned}
\end{equation}
The first equation motivates us to decompose the spinor using the projector $\Pi_{z}^{\pm} = \frac{1}{2}(1\pm \Gamma_{z})$. As a result, the collective expression of the solution is 
\begin{equation}
\epsilon_{\AdS{5}} = \left[\left(\sqrt{z} + \frac{1}{\sqrt{z}}x^{\alpha}\Gamma_{\alpha}\right)\frac{1+\Gamma_{z}}{2} + \frac{1}{\sqrt{z}}\frac{1-\Gamma_{z}}{2}\right]\epsilon_{0}\,,
\end{equation}
where $\epsilon_{0}$ is a four-component constant spinor, and all the Gamma matrices above have flat indices\,. In the whole $\AdS{5} \times S^{5}$, the expression is slightly extended to 
\begin{equation}
\epsilon_{\AdS{5}} = \left[\left(\sqrt{z} + \frac{1}{\sqrt{z}}x^{\alpha}\Gamma_{\alpha} \Gamma_{z}\right)\frac{1-i \gamma \Gamma_{z}}{2} + \frac{1}{\sqrt{z}}\frac{1+ i\gamma \Gamma_{z}}{2}\right](\epsilon_{0} + i \chi_{0})\,,
\end{equation}
where $\gamma = \Gamma_{\theta_{1}\theta_{2} \phi_{1}\phi_{2}\phi_{3}}= \Gamma_{56789}$ is a product of all the flat Gamma matrices for $S^{5}$.
\subsection{$\AdS{5}$ in Poincar\'e, light-cone, and polar coordinates}
Let us include the light-cone and polar coordinates in the previous coordinate system:
\begin{equation}
ds^{2}_{\AdS{5}} = \frac{-2dx^{+}dx^{-} + (d\rho)^{2} + \rho^{2}(d\phi)^{2} + (dz)^{2}}{z^{2}}\,.
\end{equation}
Note that the flat metric in the light-cone part has non-zero off-diagonal components:
\begin{equation}
\eta_{+-} = \eta_{-+} = -1\,,\quad \eta_{++} = \eta_{--} = 0\,.
\end{equation}  
Using the vielbein one-form 
\begin{equation}
\textbf{e}^{\pm} = \frac{dx^{\pm}}{z}\,,\quad \textbf{e}^{\rho} = \frac{d\rho}{z}\,,\quad \textbf{e}^{\phi} = \frac{\rho d\phi}{z}\,,\quad \textbf{e}^{z} = \frac{dz}{z}\,,
\end{equation}
the non-zero spin connections are given by
\begin{equation}
\omega_{\pm}^{z \pm} = \omega_{\rho}^{z \rho} = \frac{1}{z}\,,\quad \omega_{\phi}^{z \phi} = \frac{\rho}{z}\,,\quad \omega_{\phi}^{\phi \rho} = 1\,. 
\end{equation}
Then the gravitino equation can be written as 
\begin{equation}
\begin{aligned}
\partial_{\pm}\epsilon &= \frac{1}{2z}\Gamma_{\pm}(1+\Gamma_{z})\epsilon\,,\\
\partial_{\rho}\epsilon &= \frac{1}{2}\Gamma_{\rho }(1+\Gamma_{z})\epsilon\,,\\
\partial_{\phi}\epsilon &= \frac{\rho}{2z} \Gamma_{\phi}(1+\Gamma_{z})\epsilon + \frac{1}{2}\Gamma_{\rho\phi}\epsilon\,,\\
\partial_{z}\epsilon &= \frac{1}{2z}\Gamma_{z}\,.
\end{aligned}
\end{equation}
It is not hard to see that the solution takes the form
\begin{equation}
\epsilon_{\AdS{5}} = \left[\left(\sqrt{z} + \frac{1}{\sqrt{z}}(x^{+}\Gamma_{+} + x^{-}\Gamma_{-} + \rho\Gamma_{\rho})\right)e^{\frac{\phi}{2}\Gamma_{\rho\phi}}\frac{1+\Gamma_{z}}{2} + \frac{1}{\sqrt{z}}e^{\frac{\phi}{2}\Gamma_{\rho\phi}}\frac{1-\Gamma_{z}}{2}\right]\epsilon_{0}\,,
\end{equation}
where $\epsilon_{0}$ denotes a four-component constant spinor.
\subsection{$S^{5}$}
We parametrize the five-sphere such that the $U(1)^{3}$ symmetry is manifest. 
The metric is given by
\begin{equation}
  \begin{aligned}
  \dd{s}_{S^{5}}^{2} &= \sum_{i=1}^{3}\dd{\rho_{i}}^{2} + \rho_{2}^{2} \dd{\phi_{i}}^{2}\\
   &=
   \begin{multlined}[t]
\dd{\theta_2}^{2} + \sin[2](\theta_2) \dd{\theta_1}^{2}  + \sin[2](\theta_2)  \cos[2](\theta_1) \dd{\phi_1}^{2} \\
  + \sin[2](\theta_2) \sin[2](\theta_1) \dd{\phi_2}^{2} + \cos[2]( \theta_2) \dd{\phi_3}^{2} \,, 
\end{multlined}
\end{aligned}
\end{equation}
where the three $\rho_{i}$'s were parametrized in the second step as  
\begin{equation}
\rho_{1} = \cos(\theta_1)\sin(\theta_2)\,,\quad \rho_{2} = \sin(\theta_1) \sin(\theta_2)\,,\quad \rho_{3} = \cos(\theta_2)\,.
\end{equation}
A natural coframe is defined as
\begin{equation}
\begin{aligned}
\mathbf{e}^{\theta_{1}} &= \sin(\theta_2) \dd{\theta_{1}}\,, & \mathbf{e}^{\theta_{2}} &= \dd{\theta_{2}}\,,\\
\mathbf{e}^{\phi_{1}} &= \cos(\theta_1)\sin(\theta_2) \dd{\phi_{1}}\,, & \mathbf{e}^{\phi_{2}} &= \sin(\theta_1)\sin(\theta_2) \dd{\phi_{2}}\,,\quad \mathbf{e}^{\phi_{3}} = \cos(\theta_2) \dd{\phi_{3}}\,,
\end{aligned} 
\end{equation}
leading to the spin connections
\begin{equation}
\begin{aligned}
\omega^{\theta_{1}\theta_{2}}_{\theta_{1}} &= \cos(\theta_2)\,, & \omega^{\theta_{1}\phi_{1}}_{\phi_{1}} &= \sin(\theta_1)\,, & \omega^{\phi_{2}\theta_{1}}_{\phi_{2}} &= \cos(\theta_1)\,,\\
\omega^{\phi_{1}\theta_{2}}_{\phi_{1}} &= \cos(\theta_1)\cos(\theta_2)\,,& \omega^{\phi_{2}\theta_{2}}_{\phi_{2}} &= \sin(\theta_1)\cos(\theta_2)\,, & \omega^{\theta_{2}\phi_{3}}_{\phi_{3}} &= \sin(\theta_2)\,.
\end{aligned}
\end{equation}
Hence, the $S^{5}$ part of the gravitino equation has the following components:
\begin{equation}
\begin{aligned}
\partial_{\theta_{1}} \epsilon + \frac{1}{2}\cos(\theta_2)\Gamma_{\theta_{1}\theta_{2}}\epsilon &=\frac{i}{2}\sin(\theta_2)\Gamma_{\theta_{1}}\epsilon\,,\\
\partial_{\theta_{2}}\epsilon &= \frac{i}{2}\Gamma_{\theta_{2}}\epsilon\,,\\
\partial_{\phi_{1}}\epsilon + \frac{1}{2}\left(\sin(\theta_1)\Gamma_{\theta_{1}\phi_{1}}+ \cos(\theta_1)\cos(\theta_2)\Gamma_{\phi_{1}\theta_{2}}\right)\epsilon &= \frac{i}{2}\cos(\theta_1)\sin(\theta_2)\Gamma_{\phi_{1}}\epsilon\,,\\
\partial_{\phi_{2}}\epsilon + \frac{1}{2}\left(\cos(\theta_1)\Gamma_{\phi_{2}\theta_{1}}+\sin(\theta_1)\cos(\theta_2)\Gamma_{\phi_{2}\theta_{2}}\right)\epsilon &=\frac{i}{2}\sin(\theta_1)\sin(\theta_2)\Gamma_{\phi_{2}}\epsilon\,,\\
\partial_{\phi_{3}}\epsilon + \frac{1}{2}\sin(\theta_2)\Gamma_{\theta_{2}\phi_{3}}\epsilon &= \frac{i}{2}\cos(\theta_2)\Gamma_{\phi_{3}}\epsilon.
\end{aligned}
\end{equation}
Consequently, the Killing spinor on $S^{5}$ only is given by
\begin{equation}
\label{eq:spinor-s5}
\epsilon_{S^{5}} = e^{\frac{i}{2}\theta_{2}\Gamma_{\theta_{2}}}e^{\frac{i}{2}\phi_{3}\Gamma_{\phi_{3}}}e^{\frac{\theta_{1}}{2}\Gamma_{\theta_{2}\theta_{1}}}e^{\frac{\phi_{1}}{2}\Gamma_{\theta_{2}\phi_{1}}}e^{\frac{\phi_{2}}{2}\Gamma_{\theta_{1}\phi_{2}}}\tilde{\epsilon}_{0}\,,
\end{equation}
where $\tilde{\epsilon}_{0}$ is {a constant spinor}, and all the Gamma matrices are flat. 
\section{Expression for the complexified Killing spinor }
It is convenient to define the complexified Killing spinor in order to calculate the supersymmetry variations in  type IIB backgrounds for a consistency check. Let us define
\begin{equation}
\epsilon_{\mathbb{C}} \equiv \epsilon_{+} + i \epsilon_{-}\,.
\end{equation}
The the dilatino and gravitino variations can be rewritten in terms of $\epsilon_{\mathbb{C}}^{(\ast)}$ as follows:
\begin{equation}
\begin{aligned}
\delta \lambda_{+}+i\delta \lambda_{-} &= \frac{1}{2}\slashed{\partial}\Phi\epsilon_{\mathbb{C}} + \frac{1}{2}(I^{m}B_{mn}\Gamma^{n}\epsilon_{\mathbb{C}} + I_{n}\Gamma^{n}\epsilon_{\mathbb{C}}^{\ast})- \frac{1}{24}\slashed{H}_{3}\epsilon_{\mathbb{C}}^{\ast} - \frac{i}{8}\left[\slashed{\mathcal{F}}_{1}\epsilon_{\mathbb{C}} - \frac{1}{12}\slashed{\mathcal{F}}_{3} \epsilon_{\mathbb{C}}^{\ast}\right]\\
\delta \Psi_{+m}+ i \delta \Psi_{-m} &= \nabla_{m}\epsilon_{\mathbb{C}} - \frac{1}{8}H_{mnp}\Gamma^{np}\epsilon_{\mathbb{C}}^{\ast}  - \frac{i}{8\cdot 6}\slashed{\mathcal{F}}_{3}\Gamma_{m}\epsilon_{\mathbb{C}}^{\ast} + \frac{i}{8}\left[\slashed{\mathcal{F}}_{1} + \frac{1}{240} \slashed{\mathcal{F}}_{5}\right]\Gamma_{m}\epsilon_{\mathbb{C}}\,.
\end{aligned}
\end{equation}
It is clear that the action of the projector changes in this notation. For example, if we impose
\begin{equation}
\Pi\, \epsilon  = \frac{1}{2}(1-\Gamma_{0123} \otimes (i\sigma_{2}))\epsilon =0\,,
\end{equation}
then we find for the complexified Killing spinor
\begin{equation}
\Pi^{\prime}\epsilon_{\mathbb{C}} = \frac{1}{2}(1+i\Gamma_{0123})\epsilon_{\mathbb{C}}=0\,.
\end{equation} 

\def\baselinestretch{1}

\begin{small}
  \dosserif
  \bibliography{bibliography}{}
  \bibliographystyle{JHEP}
\end{small}

\end{document}